\documentclass[preprint,showpacs,showkeys]{revtex4}

\usepackage{amssymb, amsmath, amsfonts, latexsym, verbatim, mathrsfs}

\usepackage{graphicx}

\newtheorem{itlemma}{Lemma}[section]
\newtheorem{itproposition}[itlemma]{Proposition}
\newtheorem{itcorollary}[itlemma]{Corollary}
\newtheorem{itremark}[itlemma]{Remark}
\newtheorem{itremarks}[itlemma]{Remarks}
\newtheorem{itdefinition}{Definition}
\newtheorem{itexample}[itlemma]{Example}

\newenvironment{remark}{\begin{itremark}\rm}{\end{itremark}} 

\newenvironment{proposition}{\begin{itproposition}\rm}{\end{itproposition}}
\newenvironment{definition}{\begin{itdefinition}\rm}{\end{itdefinition}}

\begin{document}

\title{Geometric analysis of minimum time trajectories for a two-level quantum system}
\thanks{I am grateful to D. D'Alessandro for many helpful discussions.
Work supported by ARO MURI under Grant W911NF-11-1-0268}

\author{Raffaele Romano}
\email{rromano@iastate.edu}
\affiliation{Department of Mathematics, Iowa State University, Ames, IA, USA}

%


\begin{abstract}
We consider the problem of controlling in minimum time a two-level quantum
system which can be subject to a drift. The control is assumed to be bounded in magnitude,
and to affect two or three independent generators of the dynamics.
We describe the time optimal trajectories in $SU(2)$, the Lie group of possible evolutions for the system,
by means of a particularly simple parametrization of the group. A key ingredient of our analysis is the
introduction of the {\it optimal front line}.
This tool allows us to fully characterize the time-evolution of the reachable sets, and to derive
the worst-case operators and the corresponding times. The analysis is performed in any regime: controlled dynamics stronger,
of the same magnitude or weaker than the drift term, and gives a method to synthesize quantum logic
operations on a two-level system in minimum time.
\end{abstract}



\maketitle


\section{Introduction}

Control theory studies how the dynamics of a system can be modified through suitable
external actions called {\it controls}~\cite{miko}. When applied to quantum systems, it provides
tools for the study of the feasibility and optimization of particular operations, for
instance, in quantum information processing~\cite{nielsen}, in atomic and molecular physics,
and in Nuclear Magnetic Resonance (NMR)~\cite{levitt}. In this work, we explore the time-optimal
control~\cite{miko2,khaneja} of the dynamics of a two-level system (or qubit), the basic unit in quantum information and
quantum computation. For its fundamental role, the control of this system
has been studied in several works, under many different assumptions (see for example~\cite{wu,boscain,wenin,kirillova,garon,hegerfeldt,albertini} and
references therein). Here, we provide a complete
characterization of the time-optimal trajectories, assuming that the dynamics can contain a
non-controllable part (the drift), and that the controllable part depends on two or three independent control
functions.

From a mathematical point of view, we introduce some new key tools which enable a
simple and comprehensive treatment of the system, and the extension of the results in~\cite{albertini}.
In particular, our analysis holds for any relative strength between controllable and non-controllable
dynamics. The drift might be a dominant contribution, a perturbation, or a comparable term with
respect to the controlled part. Our analysis is relevant whenever it is
not accurate to assume that quantum operations can be performed in null time, that is, through
infinitely strong controls.

\noindent The system dynamics is expressed through the Schr\"{o}dinger operator equation
\begin{equation}\label{scro}
    \dot{X}(t) = -i (\omega_0 S_z + u_x S_x + u_y S_y + u_z S_z) X(t)
\end{equation}
with initial condition $X(0) = I$. The operator $X(t)$, an element of the special unitary
group $SU(2)$, realizes the time evolution as $\rho(t) = X(t) \rho(0) X^{\dagger}(t)$,
where $\rho(t)$ is the statistical operator associated to the system. The three functions of
time $u_k = u_k (t)$, $k = x, y, z$ are the control parameters, which we assume bounded by
\begin{equation}\label{contbound}
    u_x^2 + u_y^2 + u_z^2 \leq \gamma^2.
\end{equation}
Later, we will assume that only $u_x$ and $u_y$ can be used to
affect the dynamics, i. e., we will set $u_z = 0$. The generators $S_k$, $k = x, y, z$,
are given by $S_k = \frac{1}{2} \sigma_k$, where
\begin{equation}\label{pauli}
    \sigma_x = \left(
           \begin{array}{cc}
           0 & 1 \\
           1 & 0 \\
           \end{array}
           \right), \quad
     \sigma_y = \left(
            \begin{array}{cc}
            0 & -i \\
            i & 0 \\
            \end{array}
            \right), \quad
     \sigma_z = \left(
             \begin{array}{cc}
             1 & 0 \\
             0 & -1 \\
             \end{array}
             \right) \quad
\end{equation}
are the Pauli matrices, with commuting relations $[\sigma_k, \sigma_l] = 2 i \sigma_m$,
where $(k, l, m)$ is a cyclic permutation of $(x, y, z)$. The first
contribution might be an arbitrary, static drift term which
can always be written as in (\ref{scro}) by a suitable redefinition of the
Pauli matrices and of the control functions.

In this work, we characterize the time optimal trajectories in $SU(2)$ for any
final operator $X_f = X (t_f)$, where $t_f$ is the minimum
time for the transition $I \rightarrow X_f$. We derive the corresponding
optimal controls $u_x$, $u_y$ and $u_z$ for arbitrary values of $\omega_0$
and $\gamma$, and provide a complete description of the reachable sets in $SU(2)$
at any time $t$, that is, the family of operators the system evolution can be mapped to
in the given time $t$ (see Definition \ref{def2} below).
In particular, we derive worst-case operators and times.
It follows from standard results in geometric control theory that the system is controllable,
and every final $X_f$ can be reached at some finite time.
An important ingredient of our analysis is a representation of
elements of the special unitary group which solely relies on {\it two} parameters,
providing a clear description of optimal trajectories and reachable sets
in terms of the evolution of the boundary of the reachable sets themselves.

The plan of this work is as follows. We start our analysis by considering a system with three arbitrary
controls (in control theoretical jargon, a {\it fully actuated system}). In Section~\ref{Sec1} we review the
Pontryagin maximum principle of optimal control~\cite{pontryagin}, which is the starting point of our analysis,
and we derive the necessary conditions for optimal controls. By using them, in Section~\ref{Sec2}
we explicitly compute the candidate optimal trajectories in $SU(2)$, and represent them
in the chosen parametrization of the special unitary group. We introduce the notion
of {\it optimal front line}, which describes the evolution of the boundary of the
reachable set. By using it, in Section~\ref{Sec3} we characterize the evolution of
reachable sets in the three cases $\gamma > \vert \omega_0 \vert$, $\gamma = \vert \omega_0 \vert$ and
$\gamma < \vert \omega_0 \vert$, and provide the optimal times, whenever an
analytical expression is possible. We also derive the worst-case operators
and the relative times. In  Section~\ref{Sec4} we use the same ideas and
formalism to fully characterize the reachable sets (and related quantities) in the case where only two
controls affect the dynamics. In such a case, we find that there are different evolution for the reachable
sets in the cases $\gamma \geqslant \vert \omega_0 \vert$, $\frac{1}{\sqrt{3}} \vert \omega_0 \vert < \gamma < \vert \omega_0 \vert$
and $\gamma \leqslant \frac{1}{\sqrt{3}} \vert \omega_0 \vert$.
In Section~\ref{Sec5} we provide examples of applications by particularizing our results to some special
target operators: diagonal operators and the SWAP operator. This is done in
both scenarios of two of three controls. In Section~\ref{Sec6} we compare our work to existing results on the
optimal control on $SU(2)$, describe possible extensions of the approach, and finally conclude.

\section{The Pontryagin maximum principle}\label{Sec1}

Given an arbitrary final operator $X_f$, we consider a trajectory $X(t)$ in
$SU(2)$, determined by control functions $v_k = v_k(t)$, $k = x, y, z$, such that
$X(0) = I$ and $X(t_f) = X_f$.
A basic tool for the study of optimal control problems is given by the
Pontryagin maximum principle, which, in the context of control of system (\ref{scro}) on the
Lie group $SU(2)$, takes the following form.
\begin{definition}\label{def1}
The {\it Pontryagin Hamiltonian} is defined as
\begin{equation}\label{pontha}
    H(M, X, v_x, v_y, v_z) = i \Big( \omega_0 \langle M, X^{\dagger} S_z X \rangle +
    \sum_{k = x, y, z} v_k \langle M, X^{\dagger} S_k X \rangle \Big)
\end{equation}
where $M \in \mathfrak{su}(2)$, and
$\langle A, B \rangle \equiv {\rm Tr} \, (A^{\dagger} B)$.
\end{definition}\label{def2}
\begin{proposition}
({\it Pontryagin maximum principle})
Assume that a control strategy $u_k = u_k(t)$, $k = x, y, z$, with $u_x^2 + u_y^2 + u_z^2 \leqslant \gamma^2$,
and the corresponding trajectory
$\tilde{X}(t)$ are optimal (that is, the final time $t_f$ is minimal). Then there exists $\tilde{M} \in \mathfrak{su}(2)$, $\tilde{M} \ne 0$, such that $H(\tilde{M}, \tilde{X}, u_x, u_y, u_z) \geqslant H(\tilde{M}, \tilde{X}, v_x, v_y, v_z)$ for every $v_x, v_y, v_z$ such that $v_x^2 + v_y^2 + v_z^2 \leqslant \gamma^2$. 
\end{proposition}
Define the coefficients
\begin{equation}\label{bdef}
    b_k = i \langle M, X^{\dagger} S_k X \rangle, \quad k = x, y, z.
\end{equation}
By using the Lagrange multipliers method to maximize (\ref{pontha}) with
the bound (\ref{contbound}), we find that the optimal controls satisfy
\begin{equation}\label{optcont}
    u_k = \gamma \frac{b_k}{\sqrt{b_x^2 + b_y^2 + b_z^2}}, \quad k = x, y, z.
\end{equation}
Arcs where the Pontryagin Hamiltonian is independent of the control functions are called {\it singular} and, on them,
the controls are not constrained by equations like (\ref{optcont}). In general, an optimal trajectory
will be a concatenation of singular and non-singular arcs, and, usually, the presence of singular arcs makes the
solution of the optimal control problem more difficult, and more sophisticated mathematical tools are required
to face the problem (see for instance~\cite{boscain2} for a general analysis on 2-dimensional manifolds, or~\cite{wu2,lapert}
for some recent applications of the Pontryagin maximum principle when singular arcs are present). In our scenario, a singular
trajectory would require $b_x = b_y = b_z = 0$ in some interval, but this is impossible, because
in this case 
$\tilde{M}$ would vanish, and this is excluded by the Pontryagin principle. Therefore, we can conclude that trajectories
containing singular arcs are never optimal.

The dynamics of the $b_k$ coefficients can be derived by differentiating (\ref{bdef}) with respect to $t$,
and by using the commutation relations among Pauli matrices. We find
\begin{eqnarray}\label{bdyn}
  \dot{b}_x &=& - (\omega_0 + u_z) b_y + u_y b_z, \nonumber \\
  \dot{b}_y &=& (\omega_0 + u_z) b_x - u_x b_z, \\
  \dot{b}_z &=& u_x b_y - u_y b_x. \nonumber
\end{eqnarray}
By considering the form of optimal controls (\ref{optcont}), we obtain that $\dot{b}_z = 0$,
that is, $b_z$ is constant. Moreover, using (\ref{optcont}) in (\ref{bdyn}),
\begin{equation}\label{bxby}
  b_x = \mu \cos{(\omega_0 t + \varphi)}, \quad
  b_y = \mu \sin{(\omega_0 t + \varphi)},
\end{equation}
where $\mu \geq 0$ and $\varphi$ are constants. Therefore, the candidate optimal controls are given by
\begin{eqnarray}\label{optcont2}
  u_x &=& \gamma \sqrt{1 - \alpha^2} \cos{(\omega_0 t + \varphi)} \nonumber \\
  u_y &=& \gamma \sqrt{1 - \alpha^2} \sin{(\omega_0 t + \varphi)} \\
  u_z &=& \gamma \alpha, \nonumber
\end{eqnarray}
and $\alpha \in [-1, 1]$ is given by
\begin{equation}\label{alpha}
    \alpha = \frac{b_z}{\sqrt{b_x^2 + b_y^2 + b_z^2}}.
\end{equation}
Because of the special form of the candidate optimal controls, the dynamics (\ref{scro}) can be integrated,
as we prove in the next section.

\section{Extremal trajectories in SU(2)}\label{Sec2}

We substitute the extremals~\footnote{Candidate optimal controls and trajectories are called extremals in optimal
control theory language.} in (\ref{scro}), and find the
corresponding extremal trajectories in $SU(2)$. To proceed, it is convenient to
counter-evolve the drift of the system, by passing to the interaction picture of the dynamics,
\begin{equation}\label{Z}
    Z = e^{i \omega_0 S_z t} X, \quad Z(0) = I.
\end{equation}
With this substitution the differential evolution for $Z$ is given by
\begin{equation}\label{scroZ}
    \dot{Z} = -i \gamma \Big( \sqrt{1 - \alpha^2} (\cos{\varphi} \,
    S_x + \sin{\varphi} \, S_y) + \alpha S_z \Big) Z,
\end{equation}
which is simpler to integrate because the generator is time-independent. In the
adopted representation we find that
\begin{equation}\label{Zfin}
    Z = \left(
          \begin{array}{cc}
            \cos{\gamma \tau} - i \alpha \sin{\gamma \tau} & -i
            \sqrt{1 - \alpha^2} e^{-i \varphi} \sin{\gamma \tau} \\
            -i \sqrt{1 - \alpha^2} e^{i \varphi} \sin{\gamma \tau} &
            \cos{\gamma \tau} + i \alpha \sin{\gamma \tau} \\
          \end{array}
        \right),
\end{equation}
where we have defined $\tau = \frac{t}{2}$ to simplify the notation. Throughout this paper, we will switch between $t$
and the re-scaled time $\tau$, possibly with subscripts, without further comments. By using (\ref{Z}), we compute
\begin{equation}\label{Xfin}
    X = \left(
          \begin{array}{cc}
            e^{-i \omega_o \tau} \Big(\cos{\gamma \tau} - i \alpha \sin{\gamma \tau}\Big) & -i
            \sqrt{1 - \alpha^2} e^{-i (\omega_o \tau + \varphi)} \sin{\gamma \tau} \\
            -i \sqrt{1 - \alpha^2} e^{i (\omega_o \tau + \varphi)} \sin{\gamma \tau} &
            e^{i \omega_o \tau} \Big(\cos{\gamma \tau} + i \alpha \sin{\gamma \tau} \Big) \\
          \end{array}
        \right).
\end{equation}
This is the form of the extremal trajectories in $SU(2)$. They depend on the two parameters $\alpha$
and $\varphi$ (which can be tuned via $u_x$, $u_y$, $u_z$) as well as on the fixed parameters $\omega_0$ and $\gamma$.
To find the optimal trajectory for a given final state $X_f$, one has to determine the values of
$\alpha$ and $\varphi$ such that the transition $I \rightarrow X_f$ takes the minimal time.
This is conveniently done by choosing a suitable representation of $SU(2)$, described in the following.

\begin{remark}\label{rapper}
An arbitrary operator $X \in SU(2)$ can be given the following representation:
\begin{equation}\label{su2}
    X = \left(
          \begin{array}{cc}
            r e^{i \psi} & \sqrt{1 - r^2} e^{i \phi} \\
            -\sqrt{1 - r^2} e^{-i \phi} & r e^{-i \psi} \\
          \end{array}
        \right).
\end{equation}
Therefore, $X$ is described in terms of three parameters: $r$, $\psi$ and $\phi$.
It turns out that, in the control scenario at hand, the optimal time does not
depend on the parameter $\phi$. In other words, all the operators in $SU(2)$
which differ only for the value on $\phi$ are reached in the same optimal time.
In fact, in (\ref{Xfin}) it is possible to arbitrarily change
the phase of the off-diagonal terms by suitably choosing $\varphi$. This parameter
enters the analysis only in the phase of the off-diagonal terms, and it is independent
of the choice of $\alpha$. Therefore, to fully characterize the optimal trajectories
in $SU(2)$ and the reachable sets, we can limit our attention to the upper diagonal
element of $X(t)$, which is sufficient to determine $r$ and $\psi$.

This result can also be proven by adopting the argument in Proposition 2.1 in~\cite{albertini}, where it
is shown that the minimum time to reach $X$ and $e^{i \xi \sigma_z} X e^{-i \xi \sigma_z}$ is the same
for all real $\xi$ when there are two controls $u_x$ and $u_y$. The two operators differ only for the
phase of the off-diagonal entries. The proof is valid also for a fully actuated system.
\end{remark}
According to the previous remark, we shall parameterize $SU(2)$ solely by $r$ and $\psi$, or $x$ and $y$
in the equivalent representation $r e^{i \psi} = x + i y$. A point in the unit disk in the $(x, y)$ plane
represents a family of matrices in $SU (2)$ which only differ by the phase of the anti-diagonal elements.
These matrices are reached in the same minimum time. Moreover, every candidate optimal trajectory
can be represented by its projection onto the unit disk with the understanding that any trajectory
corresponds to a family of trajectories only differing by the phase $\varphi$. Points on the
border of the unit disk ($r = 1$) correspond to diagonal matrices, and the initial point,
the identity matrix, corresponds to the point $x = 1$, $y = 0$.

\noindent By direct inspection of (\ref{Xfin}) we have
\begin{eqnarray}\label{xy}
  x_{\alpha} (\tau) &\equiv& x (\tau) = \cos{\omega_o \tau} \cos{\gamma \tau} -
  \alpha \sin{\omega_o \tau} \sin{\gamma \tau}, \nonumber \\
  y_{\alpha} (\tau) &\equiv& y (\tau) = - \sin{\omega_o \tau} \cos{\gamma \tau} -
  \alpha \cos{\omega_o \tau} \sin{\gamma \tau}.
\end{eqnarray}
with $\alpha \in [-1, 1]$. For $\alpha = -1$ we find
\begin{equation}\label{al-1}
    x_{-1}(\tau) = \cos{(\gamma- \omega_0) \tau}, \quad y_{-1}(\tau) = \sin{(\gamma - \omega_0) \tau},
\end{equation}
and for $\alpha = 1$
\begin{equation}\label{al+1}
    x_1(\tau) = \cos{(\gamma+ \omega_0) \tau}, \quad y_1(\tau) = - \sin{(\gamma+ \omega_0) \tau}.
\end{equation}
These trajectories lie on the border of the unit disk. Moreover, by multiplying the
first equation in (\ref{xy}) by $\cos{\omega_o \tau}$, and the second
equation by $\sin{\omega_o \tau}$, and subtracting the results, we eliminate the parameter $\alpha$,
and obtain
\begin{equation}\label{reachset}
    y(\tau) \sin{\omega_o \tau} - x(\tau) \cos{\omega_o \tau} + \cos{\gamma \tau} = 0.
\end{equation}
This relation is a constraint on the terminal points of the candidate optimal trajectories at time $t = 2 \tau$, for
arbitrary $\alpha$: they lie on a line with time-dependent slope and intercept. This can also be
seen by noticing that we can recast (\ref{xy}) in the form
\begin{equation}\label{xyrecast}
	x_{\alpha}(\tau) + i y_{\alpha}(\tau) = \frac{1 - \alpha}{2} \, e^{i (\gamma - \omega_0) \tau} +
	\frac{1 + \alpha}{2} \, e^{-i (\gamma + \omega_0) \tau},
\end{equation}
which explicitly shows the special role played by the trajectories with $\vert \alpha \vert = 1$.

As $\alpha$ varies in $[-1, 1]$, Eq.~(\ref{xyrecast}) describes a segment connecting two points on the unit circle.
The end points rotate with uniform speed on the disk border, unless
$\vert \gamma \vert = \omega_0$, since in this case one of them is fixed in $(1, 0)$.
Extremal trajectories are parameterized by $\alpha \in [-1, 1]$. In general, since there
is a one-to-one correspondence between points of the segment and $\alpha \in [-1, 1]$,
two extremal trajectories cannot reach the same point in exactly the same
time. Consequently, there are not overlaps points where two (or more) different
extremal trajectories intersect. The only exception to this behavior is when
the aforementioned segment collapses to a point. This scenario will arise only
when $\gamma > \omega_0$, at the worst case time.

The segment we have just described, and its generalization to the case of two controls only in Sec. \ref{Sec4},
will be a fundamental ingredient in the analysis of
reachable sets and optimal times. Therefore, we find it convenient to assign a specific
name to it: the {\it optimal front-line} ${\cal F}_t$. More precisely, we can write
${\cal F}_t \equiv {\cal F}_t (-1 \leqslant \alpha \leqslant 1)$ and we will use this notation
to represent subsets of the front line, as for instance ${\cal F}_t (\alpha_1 \leqslant \alpha \leqslant \alpha_2)$.

Given an arbitrary final state $X_f$, represented in the unit disk by $r_f e^{i \psi_f} = x_f + i y_f$,
in order to find the optimal trajectory leading to it, we have to require that $x_{\alpha} ({\tau}_f) = x_f$ and
$y_{\alpha} (\tau_f) = y_f$. The minimal time $t_f = 2 \tau_f$ is the smallest $t_f$ such that $(x_f, y_f)$ is in
the optimal front line. The corresponding $\alpha$ determines the optimal control strategy. The optimal
minimum time can also be calculated analytically or numerically as follows. From (\ref{xy}) we
find that
\begin{eqnarray}\label{reachabxy}
	r_f \cos{\psi_f} &=& \cos{\omega_0 \tau_f} \cos{\gamma \tau_f} - \alpha \sin{\gamma_0 \tau_f} \sin{\gamma \tau_f}, \nonumber \\
	r_f \sin{\psi_f} &=& - \sin{\omega_0 \tau_f} \cos{\gamma \tau_f} - \alpha \cos{\gamma_0 \tau_f} \sin{\gamma \tau_f}.
\end{eqnarray}
In (\ref{reachabxy}), if we multiply the first equation by $\cos{\omega_0 \tau_f}$, the second by $\sin{\omega_0 \tau_f}$, and
then we subtract them, we obtain
\begin{equation}\label{optti}
r_f \cos{\Big( \omega_o \tau_f + \psi_f \Big)} = \cos{\gamma \tau_f}.
\end{equation}
The minimum time is the smallest $t_f = 2 \tau_f$ for which this equation is valid.
Furthermore, by squaring the two equations in
(\ref{reachabxy}) and summing them,  we find
\begin{equation}\label{optal}
	r_f^2 = \cos^2{\gamma \tau_f} + \alpha^2 \sin^2{\gamma \tau_f},
\end{equation}
from which $\alpha$ can be found, given the prior knowledge of $t_f$. In principle, this approach can be used
to find the optimal strategy for any final target operation. However, a geometrical analysis of the optimal front line
provides much more information on how the states are reached, further insights on the optimal times, and the
geometry of the reachable sets.

\section{Properties of the reachable sets and optimal times}\label{Sec3}

\begin{definition}
The {\it reachable set at time t} is the set ${\cal R}_{t}$ of operators $Y \in SU(2)$ such that there are control
strategies $v_k$ ($k = x, y, z$), with $v_x^2 + v_y^2 + v_z^2 \leq \gamma$, driving $X(0) = I$
to $X(t) = Y$ at time $t$, under the evolution (\ref{scro}).
The {\it reachable set until time t} is the set ${\cal R}_{\leqslant t}$ of operators $Y \in SU(2)$ such that there are control
strategies $v_k$ ($k = x, y, z$), with $v_x^2 + v_y^2 + v_z^2 \leq \gamma$, driving $X(0) = I$
into $X(s) = Y$ with $s \leqslant t$, under the evolution (\ref{scro}).
\end{definition}
The two sets are related by
\begin{equation}\label{reach}
	{\cal R}_{\leqslant t} = \bigcup_{s \leqslant t} {\cal R}_{s}.
\end{equation}
The structure of these sets is a direct consequence of the evolution of the aforementioned optimal front-line ${\cal F}_t$.
It is a known fact in optimal control theory that, if a trajectory is optimal for $X_f$ at time $t$, then $X_f$ belongs
to the boundary of the reachable set until time $t$, i. e. $X_f \in \partial {\cal R}_{\leqslant t}$. Therefore,
if a point of the unit disk is reached by ${\cal F}_t$ for the first time at time $t$, it belongs to $\partial {\cal R}_{\leqslant t}$.
However, in general not all points of the optimal front line belong to $\partial {\cal R}_{\leqslant t}$,
because they might be included in front lines corresponding to earlier times. Therefore, our strategy is to
study the evolution of the front lines, and an important role will be played by the curve where ${\cal F}_t$
intersects ${\cal F}_{t + dt}$. This curve contains the points where optimal trajectories loose their optimality.
We illustrate the procedure in the three different scenarios, depending on the relative values of $\omega_0$ and $\gamma$. A generalization of this idea will be used in Sec. \ref{Sec4} as well, with the difference that
we will find several intersection curves between ${\cal F}_t$ and ${\cal F}_{t + dt}$.

\subsection{The case $\gamma > \vert \omega_0 \vert $}

This is the case where the control action is assumed to be more powerful than the natural evolution of the system.
In this case, $\gamma - \omega_0$ and $\gamma + \omega_0$ are both positive. Therefore, following (\ref{al-1}) and
(\ref{al+1}), the extremal points of the optimal front-line rotate in opposite directions along the unit-circle,
with constant angular speed. This shows that ${\cal F}_t$ and ${\cal F}_{t + dt}$ do not intersect
on the unit disk, although this fact could be proved by a direct computation. Therefore, all the trajectories
ending on the front line are optimal.
\begin{figure}[t]
  \includegraphics[width=15cm]{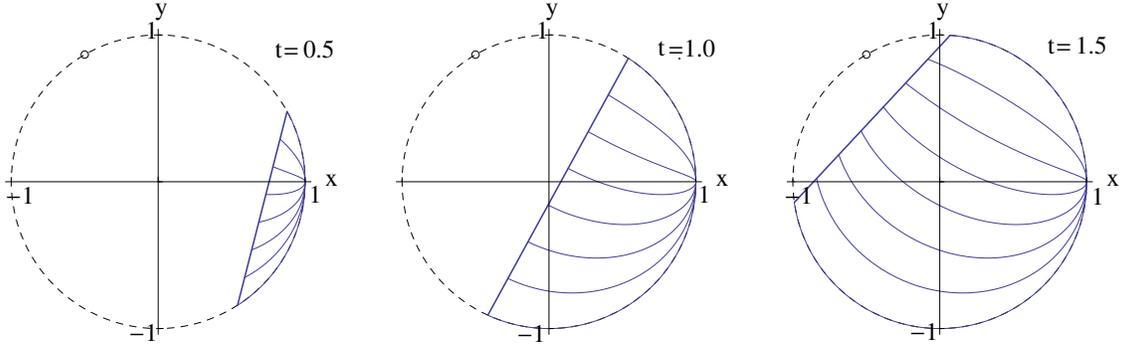}\\
  \caption{(Color online) Time evolution of the reachable sets in the unit disk, when $\gamma = 3$ and $\omega_0 = 1$. We have
  represented the optimal-front line and the optimal trajectories for several values on $\alpha$,
  at growing time. The worst-case operator is marked by a small circle.}\label{Fig1}
\end{figure}
During its evolution, ${\cal F}_t$ spans all the unit disk, and eventually collapses to a point on the border of the disk,
defined by the condition
\begin{equation}
	e^{i (\gamma - \omega_0) \tau} = e^{-i (\gamma + \omega_0) \tau}.
\end{equation}
The corresponding worst-case time is $t_{max} = \frac{2 \pi}{\gamma}$, and
$(-\cos{\pi\frac{\omega_0}{\gamma}},\sin{\pi\frac{\omega_0}{\gamma}})$ is the collapsing point. The
worst-case time is independent of $\omega_0$ because the relative angular velocity
between the extremal points of the optimal front-line depends only on $\gamma$.
Notice that $I$ and the worst-case operator are conjugate points, since there is
a one-parameter family of geodesics connecting them (the parameter is $\alpha$).

All the points
in the unit disk are reached in an optimal time $t_f \leqslant t_{max}$. See Fig.~\ref{Fig1}
for a graphical representation of the evolution of the reachable sets in a specific case.
Notice that, as a special case, we can consider $\omega_0 = 0$, that is, there is not drift in the
dynamics of the system. The corresponding worst-case operator is represented by the point $(-1, 0)$.

\subsection{The case $\gamma = \vert \omega_0 \vert$}

\begin{figure}[t]
  \includegraphics[width=15cm]{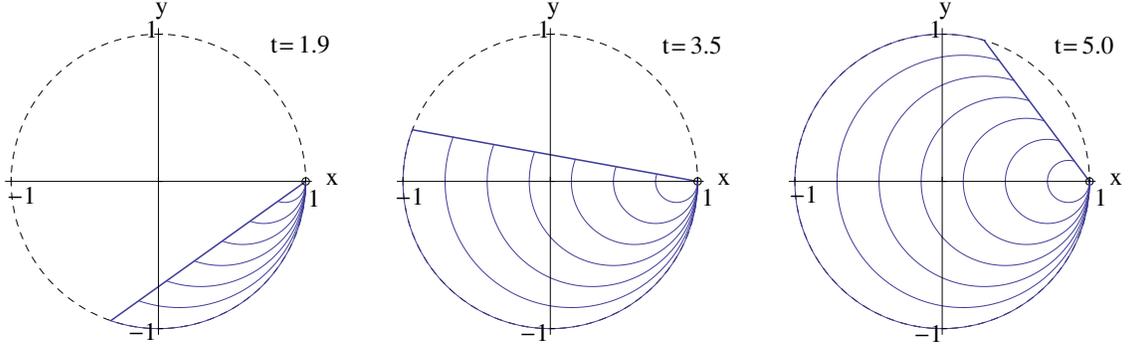}\\
  \caption{(Color online) Time evolution of the reachable sets in the unit disk, when $\gamma = 1$ and $\omega_0 = 1$. As before,
  we have represented the optimal-front line, some optimal trajectories and the worst-case operator.}\label{Fig2}
\end{figure}

In this case, the strength of the control action is the same as the free evolution of the system.
One of the extremal points of the optimal front-line is fixed at $(1,0)$, and the optimal front-line
rotates about it. This point corresponds to $\alpha = \mp 1$ when $\omega_0 = \pm \gamma$, respectively.
The analysis is analogous to the previous case, with the optimal trajectories ending on
${\cal F}_t (-1 < \alpha \leqslant 1)$ when $\omega_0 = \gamma$, and on ${\cal F}_t (-1
\leqslant \alpha < 1)$ when $\omega_0 = -\gamma$. The worst-case time is again
$t_{max}  = \frac{2 \pi}{\gamma}$, and the worst-case operator is represented by $(1, 0)$.
However, in this case it is possible to derive analytically the values for
$t_f$ and $\alpha$ for a given final state $X_f$, since the optimal trajectories are circles.
In fact, for $\omega_0 = \pm \gamma$ a direct computation shows that (\ref{xy}) is consistent with
\begin{equation}\label{circ+}
    \Big( x(\tau) - \frac{1 \mp \alpha}{2} \Big)^2 + y(\tau)^2 = \Big( \frac{1 \pm \alpha}{2} \Big)^2,
\end{equation}
and then, for any value of $\alpha$, the trajectory is a circle of radius $\frac{1 \pm \alpha}{2}$,
centered in $(\frac{1 \mp \alpha}{2}, 0)$. In the two cases,
the optimal controls for a target $X_f$, represented by $(x_f, y_f)$, are given by
\begin{equation}\label{optomga}
    \alpha = \mp \Big( x_f + \frac{y_f^2}{x_f - 1} \Big),
\end{equation}
and the optimal times are
\begin{equation}\label{opttim+}
    t_f = \frac{2 \pi}{\gamma} - \frac{1}{\gamma} \arctan{\frac{2 y_f (1 - x_f)}{y_f^2 - (1 - x_f)^2}}
\end{equation}
for $\omega_0 = \gamma$, and
\begin{equation}\label{opttim-}
    t_f = \frac{1}{\gamma} \arctan{\frac{2 y_f (1 - x_f)}{y_f^2 - (1 - x_f)^2}}
\end{equation}
for $\omega_0 = -\gamma$. See Fig~\ref{Fig2} for a pictorial representation of the evolution of the
reachable sets in a special case.

\subsection{The case $\gamma < \vert \omega_0 \vert$}

In this case, the strength of the control action is smaller than that of the free evolution. In the limit
of $\gamma$ small, the control can be seen as a perturbation to the dynamics. The analysis
is more complicated, because the optimal front-lines at time $t$ have a self-intersection during their evolution.
Therefore, some trajectories ending on the optimal front-line ${\cal F}_t$
will not be optimal. The geometric explanation of this behavior is that, in this case, the end points of ${\cal F}_t$
rotate in the same direction, generating at each time a rotation of this segment about one of its points. To
determine this point at time $t = 2 \tau$ we have to require that it is in both ${\cal F}_t$ and ${\cal F}_{t + dt}$.
\begin{figure}[t]
  \includegraphics[width=5cm]{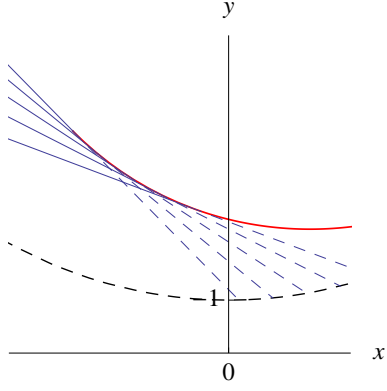}\\
  \caption{(Color online) Time evolution of the optimal front line for $\gamma = 1$ and $\omega_0 = 3$. The front lines are shown
  in even increments of time in the interval $t \in [1.2, 1.55]$. The critical optimal trajectory is the locus
  of self-intersections of these lines, where they loose their optimality. The dashed segments are the non optimal
  parts of the front lines, since these points have already been reached at a former time.}\label{criticone}
\end{figure}
According to (\ref{reachset}), we have to impose that $(x (\tau), y (\tau))$ satisfies
\begin{equation}\label{aggiunt}
    \left\{
      \begin{array}{l}
        y(\tau) \sin{\omega_o \tau} - x(\tau) \cos{\omega_o \tau} + \cos{\gamma \tau} = 0, \\
        \\
        y(\tau) \sin{\omega_o (\tau + d \tau)} - x(\tau) \cos{\omega_o (\tau + d \tau)} + \cos{\gamma (\tau + d \tau)} = 0,
      \end{array}
    \right.
\end{equation}
and the second condition can be replaced by
\begin{equation}\label{reachset2}
   y(\tau) \, \omega_0 \cos{\omega_o \tau} + x(\tau) \, \omega_0 \sin{\omega_o \tau} - \gamma
   \sin{\gamma \tau} = 0.
\end{equation}
We find that the unique solution at time $t = 2 \tau$ is given by
\begin{eqnarray}\label{solspir}
	x_c(\tau) &\equiv& \frac{\gamma}{\omega_0} \sin{\omega_o \tau} \sin{\gamma \tau}
	+ \cos{\omega_o \tau} \cos{\gamma \tau}, \nonumber \\
	y_c(\tau) &\equiv& \frac{\gamma}{\omega_0} \cos{\omega_o \tau} \sin{\gamma \tau}
	- \sin{\omega_o \tau} \cos{\gamma \tau},
\end{eqnarray}
and, by comparing (\ref{solspir}) and (\ref{xy}), we notice that
the locus of self-intersections of the optimal front-line, described by (\ref{solspir}), is itself an
extremal trajectory for the system, corresponding to $\alpha_c = - \frac{\gamma}{\omega_0}$. which we call the {\it critical trajectory}.
The value $\alpha_c$ is critical, in the sense that trajectories can be optimal
only for $\alpha \in [-1, \alpha_c]$ when $\omega_0 < 0$, and $\alpha \in [\alpha_c, 1]$ when $\omega_0 > 0$.
This can be understood by considering that for $\omega_0 < 0$ the end point of the optimal
front-line corresponding to $\alpha = - 1$ foreruns the other end point, and similarly for $\omega_0 > 0$.
Fig. \ref{criticone} shows how the optimal front lines generate the critical trajectory during their evolution.

The critical trajectory is a well-know concept in optimal geometric control theory, where it is
called the {\it cut locus}. In fact, for a given initial point, the cut locus is defined as the set of points where the extremal trajectories lose their optimality. We will shortly see that, in the regime under investigation, all the optimal trajectories lose their optimality on the critical trajectory. Therefore, our analysis of the optimal front-line represents a simple approach for determining the cut locus. Notice that, when $\gamma \geqslant \omega_0$, the cut locus reduces to a point, corresponding to the worst-case operator. This is the conjugate point to the initial point $(1, 0)$.

The critical trajectory has a singular point when $\dot{x}_c(\tau) = \dot{y}_c(\tau) = 0$. This point is a cusp singularity, whose
appearance can be geometrically understood by considering the evolution of the optimal front-line~\footnote{Generally, the optimal front line undergoes a time-dependent roto-translation in the $(x,y)$ plane. The cusp singularity appears when the translational contribution vanishes. Therefore, it represents the instantaneous rotation center of ${\cal F}_t$}. From
\begin{eqnarray}\label{losopt}
	\dot{x}_c (\tau) &=& \Big( \frac{\gamma^2 - \omega_0^2}{2 \omega_0} \Big)
	\sin{\omega_o \tau} \cos{\gamma \tau}, \nonumber \\
	\dot{y}_c (\tau) &=& \Big( \frac{\gamma^2 - \omega_0^2}{2 \omega_0} \Big)
	\cos{\omega_o \tau} \cos{\gamma \tau},
\end{eqnarray}
we find that $t_c = 2 \tau_c = \frac{\pi}{\gamma}$, and the singular point of the critical trajectory is
\begin{equation}\label{cripoi}
    x_c(\tau_c) = \frac{\gamma}{\omega_0} \sin{\frac{\pi \omega_0}{2 \gamma}},
    \quad\quad y_c(\tau_c) = \frac{\gamma}{\omega_0} \cos{\frac{\pi \omega_0}{2 \gamma}}.
\end{equation}
It turns out that this is the point where the critical trajectory looses optimality.
In fact, when $t > t_c$, the points of the critical trajectory are in the reachable
set until time $t$, and then they have already been reached at a former time.

\begin{figure}[t]
  \includegraphics[width=15cm]{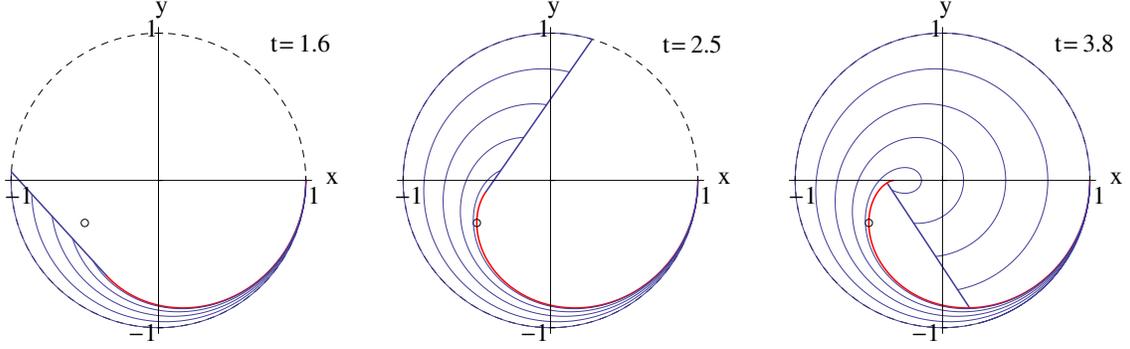}\\
  \caption{(Color online) Time evolution of the reachable sets in the unit disk, when $\gamma = 1$ and $\omega_0 = 3$. The evolution
  of the optimal-front line generates the critical optimal trajectory. The worst-case operator is marked by a small
  circle. Every trajectory loses optimality when intersecting the critical trajectory, which then represents a cut
  locus for the system.}\label{Fig3}
\end{figure}

Any other optimal trajectory looses optimality at some time, when it intersects
the reachable set until that time. The boundary of the reachable set until time $t$, $\partial
{\cal R}_{\leqslant t}$, is given by the optimal front-line and the critical trajectory. Since
the self-intersections of the optimal front line form themselves an extremal trajectory
for the system, an optimal trajectory can loose optimality only by intersecting the critical
trajectory. For this reason, as mentioned before, the critical trajectory is a cut locus for this system.

If we denote by $t_i = 4 \pi/(\gamma + \vert \omega_0 \vert)$ the time when the optimal
front-line will comes back to the point $(1,0)$, we can conclude that, for $t < {\rm Min} (t_i, t_c)$,
${\cal F}_t (-1 \leqslant \alpha \leqslant \alpha_c)$ describes the terminal points
of the optimal trajectories when $\omega_0 < 0$. Analogously, these terminal points are given by
${\cal F}_t (\alpha_c \leqslant \alpha \leqslant 1)$ when $\omega_0 > 0$.
For $t > {\rm Min} (t_i, t_c)$, the extremal trajectories which are still optimal end on
${\cal F}_t (\alpha_1 \leqslant \alpha \leqslant \alpha_2)$, where $\alpha_1$
and $\alpha_2$ are determined by the intersection of ${\cal F}_t$ and the critical
trajectory. In general, their analytical derivation is not possible. However, we
can determine the worst-case time $t_{max}$ and the corresponding $\alpha$: these are obtained by
requiring that ${\cal F}_t$ becomes tangent to the critical trajectory at some point. If we assume that
this point is reached at time $\bar{t} = 2 \bar{\tau}$, we can write it as $(x_c (\bar{\tau}), y_c (\bar{\tau}))$.
The tangent to the critical trajectory in this point is given by
\begin{equation}\label{tanopt}
	\dot{x}_c (\bar{\tau}) \Big( y - y_c(\bar{\tau}) \Big) = \dot{y}_c (\bar{\tau}) \Big( x - x_c(\bar{\tau}) \Big),
\end{equation}
which, considering the explicit expressions of $x_c$, $y_c$, $\dot{x}_c$ and $\dot{y}_c$
from (\ref{solspir}) and (\ref{losopt}), can be recast in the form
\begin{equation}\label{tanopt2}
	y \, \sin{\omega_0 \bar{\tau}} - x \,  \cos{\omega_0 \bar{\tau}} + \cos{\gamma \bar{\tau}} = 0,
\end{equation}
with $\bar{t} \in [0, t_c]$. We require that this line coincides with the optimal front-line (\ref{reachset})
at some later time $t > t_c$. Therefore
\begin{equation}\label{det1}
	\sin{\omega_0 \tau} = \pm \sin{\omega_0 \bar{\tau}}, \quad
	\cos{\omega_0 \tau} = \pm \cos{\omega_0 \bar{\tau}}, \quad
	\cos{\gamma \tau} = \pm \cos{\gamma \bar{\tau}}, \quad
\end{equation}
which, with the further constraint $0 \leqslant \bar{t} \leqslant t_c \leqslant t$, is solved by
\begin{equation}\label{pairt}
	\bar{t} = t_c \Big( 1 - \frac{\gamma}{\vert \omega_0 \vert} \Big), \qquad t = t_c \Big( 1 +
    \frac{\gamma}{\vert \omega_0 \vert} \Big).
\end{equation}
Therefore, the worst-case time for $\gamma < \vert \omega_0 \vert$ is
\begin{equation}\label{worst3}
t_{max} = 2 \tau_{max} = \frac{\pi}{\gamma} \Big( 1 + \frac{\gamma}{\vert \omega_0 \vert} \Big),
\end{equation}
which is consistent with the result found when $\gamma = \vert \omega_0 \vert$. The worst-case
point in the unit disk is arbitrarily close to $(x_c(\bar{\tau}), y_c(\bar{\tau}))$, and it is approached through the optimal trajectory characterized by $\alpha = - \alpha_c$. This can be seen by requiring that
\begin{equation}\label{sharday}
	r^2({\tau}_{max}) = r_c^2(\bar{\tau})
\end{equation}
and using that
\begin{equation}\label{tette}
	\sin{\gamma \bar{\tau}} = \sin{\gamma \tau_{max}},
\end{equation}
a direct consequence of (\ref{pairt}) and (\ref{worst3}). It turns out that (\ref{sharday}) is equivalent to
$\alpha = \alpha_c^2$, and $\alpha = \alpha_c$ is not admitted since it corresponds to the critical optimal
trajectory.
In Fig~\ref{Fig3} we provide a graphical representation of the evolution of the
reachable sets in a special case.

As $\gamma$ decreases, the critical trajectory stretches and spirals around the center of the unit disk. Eventually, when $\gamma \rightarrow 0$, the singular point of the critical trajectory approaches the center of the unit disk. In this limit, this point represents the worst case operator, which is reached only asymptotically ($t_{max} \rightarrow \infty$).

\section{The case with two controls}\label{Sec4}

In this section we consider the case where $u_z = 0$ in (\ref{scro}), that is, the
control action enters only through $S_y$ and $S_z$. This is not the most
general case of dynamics with two controls and a drift term, which could contain also
contributions along $S_y$ and $S_z$. However, the general scenario
cannot be described with the $SU (2)$ representation adopted in this work, since, in
this case, to operators differing by the phase $\phi$ in the off-diagonal elements
there usually correspond different optimal times.

\subsection{Optimal controls and trajectories}

This problem has been recently considered in~\cite{albertini} and, by using a different approach,
the optimal trajectories have been derived under the condition $\frac{1}{\sqrt{3}}
\omega_0 \leqslant \gamma \leqslant \omega_0$.Following the procedure outlined in the previous sections, we are able to fully
characterize the reachable sets (and related properties) for
arbitrary values of $\omega_0$ and $\gamma$. In particular, $\omega_0$ can be both positive, negative, or null.
Under the constraint $u_x^2 + u_y^2 \leqslant \gamma^2$, we find that
the optimal controls must satisfy
\begin{equation}\label{optcont3}
    u_k = \gamma \frac{b_k}{\sqrt{b_x^2 + b_y^2}}, \quad k = x, y
\end{equation}
and $b_k$, $k = x, y, z$ are defined as in (\ref{bdef}). Their dynamics is given by
\begin{eqnarray}\label{bdyn2}
  \dot{b}_x &=& \omega_0 b_y - u_y b_z, \nonumber \\
  \dot{b}_y &=& - \omega_0 b_x + u_x b_z, \\
  \dot{b}_z &=& -u_x b_y + u_y b_x, \nonumber
\end{eqnarray}
and, by using (\ref{optcont3}) in (\ref{bdyn2}),  we obtain that $b_z$ is constant.
Moreover, we find
\begin{equation}\label{bxby2}
  b_x = \mu \cos{(\omega t + \phi)}, \quad
  b_y = \mu \sin{(\omega t + \phi)},
\end{equation}
where $\mu$ and $\phi$ are two constants, and $\omega$ is given by
\begin{equation}\label{omega}
    \omega = \omega_0 -\frac{\gamma b_z}{\mu}.
\end{equation}
The candidate optimal controls have the form
\begin{equation}\label{optcont4}
  u_x = \gamma \cos{(\omega t + \phi)}, \qquad
  u_y = \gamma \sin{(\omega t + \phi)}.
\end{equation}
Since $b_z$ is unconstrained, $\omega$ can assume any real value.
Singular arcs are given by $b_x = b_y = 0$ on some interval, which implies $\dot{b}_z = 0$
and $u_x = u_y = 0$ in that interval. Following the argument of \cite{albertini}, it is
possible to prove that, also in this case, singular arcs can never contribute to an optimal trajectory.

Integration of the dynamics (\ref{scro}) follows the same lines outlined before
(with the intermediate operator $Z = e^{i \omega S_z t} X$), and
the final result is
\begin{equation}\label{Xfin2}
    X = \left(
          \begin{array}{cc}
            e^{-i \omega \tau} \Big(\cos{a \tau} - i \frac{b}{a} \sin{a \tau}\Big) & -i
            \frac{\gamma}{a} e^{-i (\omega \tau + \phi)} \sin{a \tau} \\
            -i \frac{\gamma}{a} e^{i (\omega \tau + \phi)} \sin{a \tau} &
            e^{i \omega \tau} \Big(\cos{a \tau} + i \frac{b}{a} \sin{a \tau} \Big) \\
          \end{array}
        \right),
\end{equation}
where we have defined $b = b (\omega) \equiv \omega_0 - \omega$, $a = a (\omega) \equiv \sqrt{b^2 + \gamma^2}$,
and $\tau = \frac{t}{2}$. The candidate optimal trajectories, in the adopted representation of $SU(2)$
(see Remark~\ref{rapper}), are obtained by taking the real and imaginary parts of the upper diagonal element in (\ref{Xfin2}):
\begin{eqnarray}\label{xy2}
  x_{\omega} (\tau) &\equiv& x (\tau) = \cos{\omega \tau} \cos{a \tau} -
  \frac{b}{a} \sin{\omega \tau} \sin{a \tau}, \nonumber \\
  y_{\omega} (\tau) &\equiv& y (\tau) = - \sin{\omega \tau} \cos{a \tau} -
  \frac{b}{a} \cos{\omega \tau} \sin{a \tau}.
\end{eqnarray}
In analogy with the case of a fully actuated system, one could numerically solve these equations for an arbitrary
final operator $X_f$ reached in minimal time $t_f = 2 \tau_f$. However, in this work we are mainly interested in
studying the evolution of the reachable sets by introducing the optimal front line and studying its evolution.

\subsection{The optimal front-line}

As before, we define the optimal front-line as the set of terminal points for a candidate optimal trajectory at time $t = 2 \tau$:
\begin{equation}\label{newofl}
	{\cal F}_t (-\infty < \omega < \infty) \equiv \{ (x_{\omega} (\tau), y_{\omega} (\tau)), -\infty < \omega < \infty\}.
\end{equation}
It is possible to verify that there is a one-to-one correspondence between $\omega$ and points on ${\cal F}_t$.
This can be seen, for instance, by rewriting (\ref{xy2}) in polar coordinates
\begin{equation}\label{unico}
	r_{\omega}^2 (\tau) = x_{\omega}^2 (\tau) + y_{\omega}^2 (\tau) = 1 - \frac{\gamma^2}{a^2} \sin^2{a \tau}
\end{equation}
and
\begin{equation}\label{unico2}
	\psi_{\omega} (t) =  \left\{
      \begin{array}{ll}
        \omega \tau + \arctan{\Big(\frac{b}{a} \tan{a \tau}\Big)},  & \quad {\rm if} \,\,0 \leqslant \tau <  \frac{\pi}{2a}, \\
        \\
        \pi + \omega \tau + \arctan{\Big( \frac{b}{a} \tan{a \tau}\Big)}, & \quad {\rm if} \,\, \frac{\pi}{2a}  < \tau <  \frac{\pi}{a}.
      \end{array}
    \right.
\end{equation}
Although it is possible to obtain $r_{\omega_1}^2 (\tau) = r_{\omega_2}^2 (\tau)$ with $\omega_1 \ne \omega_2$,
this necessarily implies $\psi_f (\omega_1) \ne \psi_f (\omega_2)$. Therefore, the correspondence
$\omega \leftrightarrow (x_{\omega} (\tau), y_{\omega} (\tau))$ is one-to-one at any $\tau$.

Not all the extremal trajectories are optimal. Following the discussion of the previous
sections, we have to consider the self-intersections of ${\cal F}_t$, as well as
the intersections of ${\cal F}_t$ and $\partial {\cal R}_{\leqslant t}$,
in order to determine critical
values of $\omega$ for which the trajectories lose optimality. In this case we must use the parametric expressions
for the points of ${\cal F}_t$ since it is not possible to solve for $\omega$ one of the two equations in (\ref{xy2}),
and obtain a closed expression of the optimal front line in terms of $x$ and $y$ alone. Therefore, we cannot directly
rely on the procedure developed in the previous sections. However, the optimal front line can be considered as the envelope
of its tangent lines.
Therefore, if there is a self-intersection of ${\cal F}_t$ in some point, there must also be a self-intersection of the tangent
line to ${\cal F}_t$ in that point. Consequently, we can find the intersections of ${\cal F}_t$ and ${\cal F}_{t + dt}$
by considering, for each $\omega$, the intersections of the tangent lines to the optimal front-line at time $t$ and $t + dt$. If they
are on ${\cal F}_t$, they correspond to the desired intersection of ${\cal F}_t$ and ${\cal F}_{t + dt}$, and
the corresponding $\omega$ is a critical value, relevant for determining where the trajectories are optimal.

Again, by means of this simple analysis we are able to fully characterize the cut loci for this system. We will find
not trivial cut loci for any value of $\omega_0$ and $\gamma$.

The slope of the tangent line to the optimal front line at time $t$, in the point labeled by $\omega$, is given by
\begin{equation}\label{sloppe}
    \frac{dy}{dx} = \frac{dy}{d\omega} \, \Big( \frac{dx}{d\omega} \Big)^{-1}.
\end{equation}
Since
\begin{eqnarray}
  \frac{dx}{d\omega} &=& -\frac{\gamma^2}{a^2} \sin{\omega \tau} (\tau \cos{a \tau} - \frac{1}{a} \sin{a \tau}), \nonumber \\
  \frac{dy}{d\omega} &=& -\frac{\gamma^2}{a^2} \cos{\omega \tau} (\tau \cos{a \tau} - \frac{1}{a} \sin{a \tau}),
\end{eqnarray}
we find that
\begin{equation}\label{sloppe2}
    \frac{dy}{dx} = \cot{\omega \tau}.
\end{equation}
Therefore, the tangent line to ${\cal F}_t$ in the point $(x_{\omega}(\tau),y_{\omega}(\tau))$, at time $t = 2 \tau$, is given by
\begin{equation}\label{tanli}
    y (\tau) \sin{\omega \tau} - x (\tau) \cos{\omega \tau} + \cos{a \tau} = 0.
\end{equation}
The intersections of tangent lines to ${\cal F}_t$ and ${\cal F}_{t + dt}$ are obtained by solving the system
\begin{equation}\label{aggiunta}
    \left\{
      \begin{array}{l}
        y(\tau) \sin{\omega \tau} - x(\tau) \cos{\omega \tau} + \cos{a \tau} = 0, \\
        \\
        y(\tau) \sin{\omega (\tau + d \tau)} - x(\tau) \cos{\omega (\tau + d \tau)} + \cos{a (\tau + d \tau)} = 0,
      \end{array}
    \right.
\end{equation}
whose solution follows the same steps which have been detailed in the previous section.
When $\omega \ne 0$, we find the unique solution~\footnote{When $\omega = 0$, the only solution to
(\ref{aggiunta}) is given by $\vert x (\tau) \vert = 1$, $y (\tau) = 0$, with the
constraint $\sin{a \tau} = 0$. This solution is already accounted for in the case $\omega \ne 0$.}
\begin{eqnarray}\label{solspir2}
	x(\tau) &=& \frac{a}{\omega} \sin{\omega \tau} \sin{a \tau}
	+ \cos{\omega \tau} \cos{a \tau}, \nonumber \\
	y(\tau) &=& \frac{a}{\omega} \cos{\omega \tau} \sin{a \tau}
	- \sin{\omega \tau} \cos{a \tau}.
\end{eqnarray}
However, since the intersection point must be on ${\cal F}_t$, also (\ref{xy2}) must be satisfied.
Therefore
\begin{eqnarray}\label{ellui}
  \frac{a}{\omega} \sin{\omega \tau} \sin{a \tau} &=& - \frac{b}{a} \sin{\omega \tau} \sin{a \tau}, \nonumber \\
  \frac{a}{\omega} \cos{\omega \tau} \sin{a \tau} &=& - \frac{b}{a} \cos{\omega \tau} \sin{a \tau},
\end{eqnarray}
which has several solutions. If $\sin{a \tau} \ne 0$,
we find that $a^2 + b \, \omega = 0$, solved by
\begin{equation}\label{cricecroc}
    \omega_c = \frac{\omega_0^2 + \gamma^2}{\omega_0}.
\end{equation}
Since this critical value is time-independent, this locus of self-intersections of the optimal front-line is by itself a critical optimal trajectory $(x_c(\tau), y_c(\tau))$. It loses its optimality at a critical time $t_c = 2 \tau_c$
such that $\dot{x}_c (\tau) = \dot{y}_c (\tau) = 0$. Since
\begin{eqnarray}\label{losopt2}
	\dot{x}_c (\tau) &=& - (\omega_0^2 + \gamma^2)
	\sin{\omega_c \tau} \cos{a \tau}, \nonumber \\
	\dot{y}_c (\tau) &=& - (\omega_0^2 + \gamma^2)
	\cos{\omega_c \tau} \cos{a \tau},
\end{eqnarray}
we find that the critical time is
\begin{equation}\label{tcri}
    t_c = \frac{\pi \vert \omega_0 \vert}{\gamma \sqrt{\omega_0^2 + \gamma^2}}.
\end{equation}
This trajectory is a cut locus for the system, analogous to that described in the case of three controls,
when $\gamma < \omega_0$.

Additional solutions to (\ref{ellui}) are found when $\sin{a \tau} = 0$. In this case the critical frequencies are implicitly
defined by $a (\omega_{c^{\prime}}) \tau = k \pi$, where $k$ is an integer. The corresponding points are on
the boundary of the unit disk: $x_{c^{\prime}} (\tau) = \cos{\omega_{c^{\prime}} \tau}$,
$y_{c^{\prime}} (\tau) = - \sin{\omega_{c^{\prime}} \tau}$. These cut loci are not optimal trajectories for the
system since the critical frequencies are time-dependent. The explicit expressions of these critical frequencies are
\begin{equation}\label{crifre2}
    \omega_{c^{\prime}}^+ (k, \tau) = \omega_0 + \sqrt{\Big(\frac{k \pi}{\tau}\Big)^2 - \gamma^2}, \quad
    \omega_{c^{\prime}}^- (k, \tau) = \omega_0 - \sqrt{\Big(\frac{k \pi}{\tau}\Big)^2 - \gamma^2},
\end{equation}
and they are defined for $\tau \leqslant \frac{k \pi}{\gamma}$, that is, $k \geqslant \frac{\gamma \tau}{\pi}$.
It turns out that $\omega_{c^{\prime}}^+ (k, \tau)  \geqslant \omega_{c^{\prime}}^- (k, \tau) $, and equality
holds only when $\tau = \frac{k \pi}{\gamma}$.
If we write $x_{c^{\prime}} (\tau)+ i y_{c^{\prime}} (\tau) = e^{i \psi_{c^{\prime}} (\tau)}$, and require that $\psi_{c^{\prime}} (0) = 0$,
we have
\begin{eqnarray}\label{spada}
  \psi_{c^{\prime}}^+ (k, \tau) &=& - \omega_{c^{\prime}}^+ (k, \tau) \tau + k \pi = - \omega_0 \tau - \sqrt{(k \pi)^2 - (\gamma \tau)^2} + k \pi, \nonumber \\
  \psi_{c^{\prime}}^- (k, \tau) &=& - \omega_{c^{\prime}}^- (k, \tau) \tau - k \pi = - \omega_0 \tau + \sqrt{(k \pi)^2 - (\gamma \tau)^2} - k \pi.
\end{eqnarray}
It is possible to prove that $\psi_{c^{\prime}}^+ (k, \tau) \geqslant \psi_{c^{\prime}}^- (k, \tau)$ for all $\tau \leqslant \frac{k \pi}{\gamma}$. Moreover,
if $k_2 > k_1$, it follows that $\psi_{c^{\prime}}^+ (k_2, \tau) < \psi_{c^{\prime}}^+ (k_1, \tau)$ and
$\psi_{c^{\prime}}^- (k_2, \tau) > \psi_{c^{\prime}}^- (k_1, \tau)$.
Since (\ref{spada}) are continuous functions, with $\psi_{c^{\prime}}^+ (k, 0) = \psi_{c^{\prime}}^- (k, 0) = 0$,
for the study of optimal trajectories we have to take $k = 1$ in (\ref{crifre2}) and (\ref{spada}). In the following,
we suppress the parameter $k$ to simplify the notation.

It turns out that, when $\omega_0 > 0$, $\psi_{c^{\prime}}^- (\tau)$ is monotonically decreasing for all $\tau$, and $\psi_{c^{\prime}}^+ (\tau)$ decreasing for $\tau < 2 \tau_c$ and increasing for $\tau > 2 \tau_c$.
Viceversa, when $\omega_0 < 0$, $\psi_{c^{\prime}}^+ (\tau)$ is monotonically
increasing for all $\tau$, and $\psi_{c^{\prime}}^- (\tau)$ increasing for $\tau < 2 \tau_c$ and decreasing for $\tau > 2 \tau_c$.
If $\omega_0 = 0$, we have $t_c = 0$, and $\psi_{c^{\prime}}^- (\tau)$, $\psi_{c^{\prime}}^+ (\tau) = - \psi_{c^{\prime}}^- (\tau)$
are monotonically decreasing and increasing for all $\tau$, respectively.
Furthermore, for $\tau \in [0, 2\tau_c)$ we have $\omega_{c^{\prime}}^- (\tau) < \omega_c < \omega_{c^{\prime}}^+ (\tau)$.
Notice that $\omega_{c^{\prime}}^+ (2 \tau_c) = \omega_c$ if $\omega_0 > 0$,  and $\omega_{c^{\prime}}^- (2 \tau_c) =
\omega_c$ if $\omega_0 < 0$.

\subsection{Reachable sets and optimal times}

The evolution of the reachable set ${\cal R}_{\leqslant t}$ is determined by the dynamics of the
optimal front line, and the treatment goes after that presented in the case of three controls.
To start with, we consider the case $\omega_0 < 0$. Following the discussion of the previous subsection,
for $t \in [0, t_c]$ the optimal trajectories at time $t$ end on ${\cal F}_t (\omega_c \leqslant \omega \leqslant \omega^+_{c^{\prime}} (\tau))$. The trajectory characterized by $\omega^+_{c^{\prime}} (\tau)$ ceases to be optimal
since it reaches the cut locus on the border of the unit disk. The other trajectories cannot intersect, since
the optimal front line has not self intersections in this range of values for $\omega$.
For $t > t_c$ the situation is more complicated. In a neighborhood of $t_c$, the optimal trajectories
are determined by ${\cal F}_t (\omega_1 \leqslant \omega \leqslant \omega^+_{c^{\prime}} (\tau))$,
where $\omega_1 > \omega_c$ can be found by intersecting the optimal front-line and the critical optimal trajectory.
The analytical expression of $\omega_1$ cannot be generically found.

The case with $\omega_0 > 0$ is analogous, and the evolution of the optimal trajectories
are described by ${\cal F}_t (\omega^-_{c^{\prime}}(\tau) \leqslant \omega \leqslant \omega_c)$ for $t \in [0, t_c]$, and by ${\cal F}_t (\omega^-_{c^{\prime}}(\tau) \leqslant \omega \leqslant \omega_2)$ for $t > t_c$,
with a suitable value $\omega_2 < \omega_c$ determined by the intersection of the optimal front-line and the
critical optimal trajectory.

\begin{figure}[t]
  \includegraphics[width=15cm]{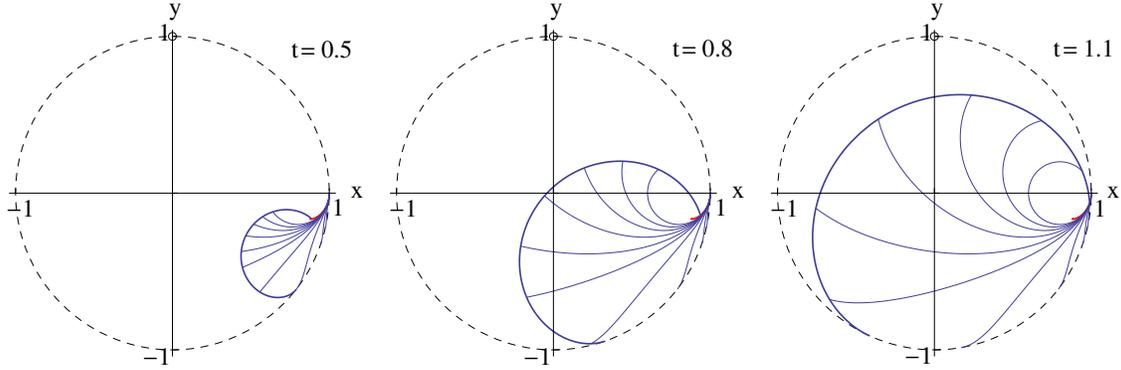}\\
  \caption{(Color online) Time evolution of the reachable sets in the unit disk, when $\gamma = 3$ and $\omega_0 = 1$. We have
  represented the optimal-front line, the critical optimal trajectory and the optimal trajectories for several values of $\omega$
  at successive times. The worst-case operator is marked by a small circle.}\label{Fig4}
\end{figure}
For the subsequent evolution, we can recognize several regimes, depending on the relative strength of controlled
and free dynamics. It is convenient to characterize where the intersection loci described by
$\omega_{c^{\prime}}^+$ and $\omega_{c^{\prime}}^-$ converge. Therefore, we impose $\psi_{c^{\prime}}^+ (\tau) - \psi_{c^{\prime}}^- (\tau) = 2 \pi$, which is solved by
\begin{equation}\label{peggiotiempo}
    t_{max} = 2 \tau = \frac{2 \pi}{\gamma}.
\end{equation}
This is the worst-time case if the point of convergence is outside the reachable set. This requirement reads $\psi_{c^{\prime}}^+ (\tau_{max}) < 2 \pi$ for $\omega_0 < 0$ or rather $\psi_{c^{\prime}}^+ (\tau_{max}) > 0$
for $\omega_0 > 0$, leading to
\begin{equation}\label{congamma1}
   \gamma \geqslant \vert \omega_0 \vert.
\end{equation}
The corresponding worst-case operator is a diagonal operator represented by $(\cos{\psi_{max}}, \sin{\psi_{max}})$, where
\begin{equation}\label{wcope}
    \psi_{max} = \psi_{c^{\prime}}^+ (\tau_{max}) = \pi \Big( 1 - \frac{\omega_0}{\gamma} \Big).
\end{equation}
It is reached along the trajectory described by $\omega = \omega_0$, which corresponds to the critical frequencies
at the final time: $\omega_{c^{\prime}}^+ (\tau_{max}) = \omega_{c^{\prime}}^- (\tau_{max}) = \omega_0$.

Notice that this analysis applies as well to the case $\omega_0 = 0$, that is, controlled dynamics without drift.
In this case the critical trajectory collapses to the initial point $(1, 0)$, which is self-conjugate.

We consider now the case $\gamma < \vert \omega_0 \vert$. The worst case state, and the corresponding time,
can be found by considering the evolution of the optimal front line. In particular, they can be determined by
requiring that ${\cal F}_t$ and the critical optimal trajectory are tangent to each other, that is, their
tangent lines overlap. In principle, this is a necessary but not sufficient condition, since there could be
several points satisfying this requirement. Nonetheless, we find that, for all $\gamma$,
there is only one possible solution. Therefore, it must correspond to the worst-case operator.
\begin{figure}[t]
  \includegraphics[width=15cm]{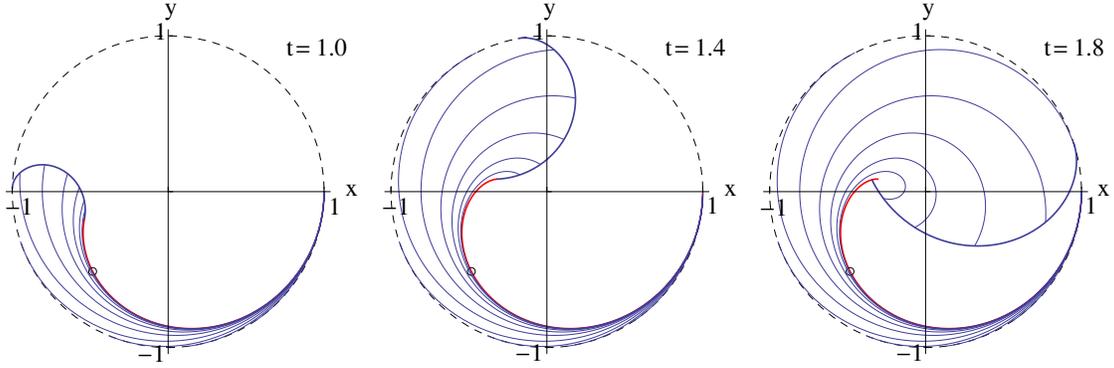}\\
  \caption{(Color online) Time evolution of the reachable sets in the unit disk, when $\gamma = 1$ and $\omega_0 = 2$.}\label{Fig5}
\end{figure}

The tangent line to the critical optimal trajectory in the point $(x_c (\bar{\tau}), y_c (\bar{\tau}))$, with $\bar{\tau} < \tau_c$, is found by using the same argument discussed in the previous section:
\begin{equation}\label{opt3cont}
    y \, \sin{\omega_c \bar{\tau}} - x \,  \cos{\omega_c \bar{\tau}} + \cos{a_c \bar{\tau}} = 0,
\end{equation}
where $a_c = a(\omega_c)$. The tangent line to ${\cal F}_t$ at a later time $t = 2 \tau > t_c$ is
given by (\ref{tanli}), and it coincides with (\ref{opt3cont}) if and only if
\begin{equation}\label{det1new}
	\sin{\omega \tau} = \pm \sin{\omega_c \bar{\tau}}, \quad
	\cos{\omega \tau} = \pm \cos{\omega_c \bar{\tau}}, \quad
	\cos{a \tau} = \pm \cos{a_c \bar{\tau}}. \quad
\end{equation}
This system of equations must be solved for $\omega$, $\tau$ and $\bar{\tau}$, with the time hierarchy
$0 \leqslant \bar{\tau} \leqslant {\tau}_c \leqslant \tau$, and $\tau$ is the
minimal time. $\tau_c$ is given by $\tau_c = \frac{t_c}{2}$ with $t_c$ as in (\ref{tcri}).
It turns out that the only possible solution to (\ref{det1new}) which satisfies the required
constraints is
\begin{equation}
	\omega \tau = \pi + \omega_c \bar{\tau}, \qquad a \tau = \pi - a_c \bar{\tau},
\end{equation}
which gives the optimal frequency
\begin{equation}\label{wcw}
	\omega = \frac{\omega_0^2 - \gamma^2}{\omega_0},
\end{equation}
and the times
\begin{eqnarray}
	\tau_{max} = \frac{\pi}{2 \vert \omega_0 \vert} \Big( 1 + \frac{\sqrt{\omega_0^2 + \gamma^2}}{\gamma} \Big), \qquad \bar{\tau} = \frac{\pi}{2 a_c}
	\frac{\vert \omega_c \vert - 2 a_c}{\vert \omega_c \vert - a_c}
\end{eqnarray}
where $a_c = a (\omega_c)$ and $\omega_c$ is given in (\ref{cricecroc}). The worst case state in the adopted representation is
arbitrary close to by the point $(x_c(\bar{\tau}), y_c(\bar{\tau}))$, approached through the optimal trajectory with $\omega$
as in (\ref{wcw}). From $\bar{\tau} \geqslant 0$, we find that these results holds for
\begin{equation}\label{congamma2}
   \gamma \leqslant \frac{1}{\sqrt{3}} \vert \omega_0 \vert.
\end{equation}

We can complete our analysis in the case
\begin{equation}\label{congamma3}
	\frac{1}{\sqrt{3}} \vert \omega_0 \vert < \gamma < \vert \omega_0 \vert
\end{equation}
by considering that the optimal front line is described by continuous functions, and then it has a smooth evolution.
It turns out that, in this regime, the optimal front line is always tangent to the critical optimal trajectory in the point $(1, 0)$, and this is the only point of intersection. Therefore, it also represents the worst case operator. The worst case time is defined by $\psi^+_{c^{\prime}} (\tau) = 2 \pi$ if $\omega_0 < 0$, or $\psi^-_{c^{\prime}} (\tau) = - 2 \pi$ if $\omega_0 > 0$, which are solved by
\begin{equation}
	\tau_{max} = \frac{2 \pi \vert \omega_0 \vert}{\omega_0^2 + \gamma^2}.
\end{equation}
When $\omega_0 < 0$ one finds $\omega^+_{c^{\prime}} (\tau_{max}) = \frac{\omega_c}{2}$, and similarly, when $\omega_0 > 0$, $\omega^-_{c^{\prime}} (\tau_{max}) = \frac{\omega_c}{2}$. Therefore, $\omega = \frac{\omega_c}{2}$ characterizes the optimal trajectory for the worst case state when (\ref{congamma3}) holds.

\section{Examples}\label{Sec5}

In this section, we derive the optimal strategy for three different target operations. This
is done for generic control strength $\gamma$ and drift $\omega_0$, in the two cases of two
or three controls affecting the dynamics.

\subsection{Diagonal operators}

Assume that the target operator is given by $X_f = e^{i \lambda \sigma_z}$, with $\lambda \in [0, 2 \pi)$ without loss of generality.
In the case of three controls, following the former discussion, we find that the optimal control strategy is given by
$\alpha = 1$ or $\alpha = -1$, depending on the relative values of $\lambda$, $\gamma$ and $\omega_0$. The optimal time is given by
\begin{equation}\label{optdiag}
    t_f = \left\{
            \begin{array}{ll}
              \frac{4 \pi - 2 \lambda}{\gamma + \omega_0}, & \hbox{if $\omega_0 \geqslant \frac{\pi - \lambda}{\pi} \gamma$} \\
              \frac{2 \lambda}{\gamma - \omega_0}, & \hbox{if $\omega_0 < \frac{\pi - \lambda}{\pi} \gamma$}
            \end{array}
          \right.
\end{equation}
and, in particular, $t_f = \frac{\lambda}{\gamma}$ if $\omega_0 = - \gamma$ or $t_f = \frac{1}{\gamma} (2 \pi - \lambda)$
if $\omega_0 = \gamma$.

In the case of two controls, the diagonal operators are always the terminal points of optimal trajectories
determined by $\omega^-_{c^{\prime}}$ or $\omega^+_{c^{\prime}}$. In general, the optimality conditions
are $\psi^+_{c^{\prime}} (\tau) = \lambda$ or $\psi^-_{c^{\prime}} (\tau) = \lambda - 2 \pi$, depending on
the specific values of $\lambda$, $\gamma$ and $\omega_0$. The optimal time is generically given by
\begin{equation}\label{optidiag2}
    t_f = \frac{2}{\omega_0^2 + \gamma^2} \Big((\pi - \lambda) \omega_0 + \Omega \Big),
\end{equation}
where $\Omega \equiv \sqrt{\pi^2 \omega_0^2 + (2 \pi \lambda - \lambda^2) \gamma^2 }$. In the specific case $\omega_0 = 0$ we obtain
\begin{equation}\label{optidiag2bis}
    t_f = \frac{2}{\gamma} \sqrt{2 \pi \lambda - \lambda^2},
\end{equation}
in accordance with the result of~\cite{garon}. The expression (\ref{optidiag2}) can be made more precise if a
specific diagonal operator is specified. For instance, if $X_f = i \sigma_z$, i.e., $\lambda = \frac{\pi}{2}$,
we find the expression
\begin{equation}\label{optidiag2ter}
    t_f = \frac{\pi}{\omega_0^2 + \gamma^2} \Big(\omega_0 + \sqrt{4 \omega_0^2 + 3 \gamma^2} \Big)
\end{equation}
which is valid for any $\omega_0$.

Following the analysis presented in Section \ref{Sec4}, we find that, when $\omega_0 > 0$,
the optimal trajectories leading to diagonal operators are characterized by $\omega \leqslant
\frac{\omega_c}{2}$, and, when $\omega_0 < 0$, by $\omega \geqslant \frac{\omega_c}{2}$. The critical
frequency $\omega_c$ has been defined in (\ref{cricecroc}). The analysis applies as well when $\omega_0 = 0$,
but in this case $\omega_c$ diverges to $\infty$ and any $\omega$ produces a trajectory ending on the
border of the unit disk. Therefore we conclude that, for $\omega_0 \ne 0$, the value $\omega =
\frac{\omega_c}{2}$ corresponds to a limit trajectory, mapping the point $(1,0)$ to itself,
and separating the trajectories leading to diagonal operators to the others trajectories, in
accordance with the result found in \cite{albertini}. This trajectory is given by
\begin{eqnarray}\label{separatrix}
    x_{\omega_c} (\tau) &=& \frac{\omega_0^2}{\omega_0^2 + \gamma^2} \cos{\frac{\omega_0^2 + \gamma^2}{\omega_0} \, \tau}
    + \frac{\gamma^2}{\omega_0^2 + \gamma^2}, \nonumber \\
    y_{\omega_c} (\tau) &=& -\frac{\omega_0^2}{\omega_0^2 + \gamma^2} \sin{\frac{\omega_0^2 + \gamma^2}{\omega_0} \, \tau},
\end{eqnarray}
therefore it is a circle of radius $\frac{\omega_0^2}{\omega_0^2 + \gamma^2}$ centered in $\Big(\frac{\gamma^2}{\omega_0^2 + \gamma^2}, 0\Big)$.

\subsection{SWAP operator}

The SWAP operator is given by $X_f = i \sigma_y$, and it is represented by $(0, 0)$,
the center of unit disk. When there are three controls, by imposing $x_{\alpha}(\tau) =
y_{\alpha}(\tau) = 0$ in (\ref{xy}), we find the optimal time $t_f = \frac{\pi}{\gamma}$ and $\alpha = 0$.
But $\alpha = 0$ is equivalent to $b_z = 0$, from the definition (\ref{alpha}) of $\alpha$.
This in turn implies $u_z = 0$ from (\ref{optcont}). Therefore, the optimal trajectories leading
to the SWAP operator are the same in the case of two or three controls, and also the optimal time is
the same. This can be directly seen by imposing $x_{\omega}(\tau) = y_{\omega}(\tau) = 0$ in (\ref{xy2}),
which is solved by $\tau = \frac{\pi}{2 a}$ and $b = 0$. This latter condition reads $\omega = \omega_0$,
and then $a = \gamma$ and $t_f = \frac{\pi}{\gamma}$, as expected. The trajectory, obtained with $\alpha = 0$
in (\ref{xy}) or rather $\omega = \omega_0$ in (\ref{xy2}), is given by
\begin{equation}\label{trajswap}
    x(\tau) + i y (\tau) = \vert \cos{\gamma \tau} \vert \,\, e^{-i \omega_0 \tau},
\end{equation}
and, in general, it does not lose optimality after reaching the SWAP operator. When $\omega_0 = 0$,
this trajectory is a segment connecting $(1, 0)$ to $(-1, 0)$, the worst-case operator.

Notice that, in the case of two controls, $\omega = \omega_0$ defines the optimal worst-case
trajectory when $\gamma > \vert \omega_0 \vert$. The worst time is twice the time to
reach the SWAP operator. Interestingly, in the same regime, the same worst time is obtained
when three controls can be used. Nonetheless, this result does not hold only for the trajectory
with $\alpha = 0$, but for any $\alpha$.

\section{Conclusions}\label{Sec6}

In this work we have studied the minimum time control of $SU(2)$ quantum operations
for a two-level quantum system. We have assumed that the system dynamics, which possibly contains
a time-independent drift term, could be externally modified by means of two or three independent
control actions. The total strength of the control is bounded, and of arbitrary magnitude when
compared to the free dynamics of the system. By using an especially simple parametrization of
the Lie group of special unitary operations, and by studying the dynamics of the boundary of
the reachable set through the evolution of the {\it optimal front line}, we are able to provide
a comprehensive description of the dynamics of the reachable sets for any relative magnitude
of the free and controlled dynamics.

Our results complement and extend former results on the behavior of optimal trajectories~\cite{albertini,garon}. We
provide a complete description of the {\it critical trajectories} in $SU(2)$, which in our context arise as loci of
self-intersections of the optimal front-line. Whenever possible, we analytically derive the optimal control strategies and the corresponding optimal times, and in each case characterize the worst-case operator and time (the so-called {\it diameter} of the system). We provide a geometrical description of the optimal control problem, which makes clear the
existence of different regimes depending on the relative strength of drift and control terms. In table~\ref{tab1} we
summarize the worst-case time in all the cases.
\begin{table}[t]
\caption{Diameter of the system in several regimes} 
\centering 
\begin{tabular}{cc cccc ccccc} 
\hline\hline 
Case & \vline & \multicolumn{3}{c}{Three controls} & \vline & \multicolumn{5}{c}{Two controls} \\ [0.5ex]
\hline 
Subcase & \vline & $\gamma < \vert \omega_0 \vert$ & \vline & $\gamma \geqslant \vert \omega_0 \vert$ & \vline & $\gamma \leqslant \frac{1}{\sqrt{3}} \vert \omega_0 \vert$ & \vline & $\frac{1}{\sqrt{3}} \vert \omega_0 \vert < \gamma < \vert \omega_0 \vert$ & \vline & $\gamma \geqslant \vert \omega_0 \vert$ \\ [1.5ex]
\hline
$t_{max}$ & \vline & $\frac{\pi}{\gamma}\Big( 1 + \frac{\gamma}{\vert \omega_0 \vert} \Big)$ & \vline & $\frac{2 \pi}{\gamma}$ & \vline & $\frac{\pi}{\vert \omega_0 \vert} \Big( 1 + \frac{\sqrt{\omega_0^2 + \gamma^2}}{\gamma} \Big)$ & \vline & $\frac{4 \pi \vert \omega_0 \vert}{\omega_0^2 + \gamma^2}$ & \vline & $\frac{2 \pi}{\gamma}$ \\[1ex] 
\hline 
\end{tabular}
\label{tab1}
\end{table}

Our results are relevant whenever quantum operations on qubits have to be engineered in the shortest possible time,
preeminently in quantum information processing, quantum communication, atomic and molecular physics, and Nuclear Magnetic Resonance.
While the case of a fully actuated system is not of primary relevance for real applications, with its simplicity
it provides the ideal framework for illustrating our technique. This approach, based on the study of the optimal front line, strongly simplifies the analysis of the system, and, when suitably generalized, it might represent a promising tool for the investigation of other optimal control problems. As an example, the present analysis of minimum time evolutions applies, with minor modifications, to the Lie group $SO (3)$, since $SU(2)$ is a double cover of it.


\end{document}